# Dissipative Rabi model for deep strong far-off-resonant driving

A.P. Saiko[1], S.A. Markevich[1] and R. Fedaruk[2]

[1]Scientific-Practical Materials Research Centre NAS of Belarus, Minsk, Belarus

[2] Institute of Physics, University of Szczecin, 70-451, Szczecin, Poland

E-mail: saiko@ifttp.bas-net.by; ryhor.fedaruk@usz.edu.pl

We study the dissipative Rabi model under deep strong far-off-resonant driving when the driving frequency and strength exceed significantly the qubit transition frequency. We find analytical expressions for the density matrix, the steady-state first-order field correlation function and the multiphoton emission spectrum. This spectrum consists of a series of narrow delta-like lines, caused by coherent processes and of Lorenzian peaks resulted from the incoherent scattering of photons. It is shown that the width of emission lines as well as
the decay rate of the time oscillations in the population difference have oscillating dependence on the driving strength. We show how the features of this dependence are controlled by the relative rates of energetic relaxation and pure dephasing.

## 1. Introduction

80 years ago Rabi [1] introduced the semi-classical model to describe the interaction of a two-level quantum system (nuclear magnetic moment) with a weak rotating magnetic field. The Rabi model is widely used for studying and control of the resonant interaction between a classical electromagnetic field and various quantum objects (atoms, nuclear and electron spins) [2, 3]. The interaction of the quantum system with a quantized electromagnetic field is considered in the frame of the quantum Rabi model. In the rotating wave approximation (RWA), where the coupling strength $g$ between the two-level system and the electromagnetic field is much smaller than the field frequency $\omega$ and the transition frequency $\varepsilon$ of the system with $|\varepsilon - \omega| \ll \varepsilon, \omega$, this model is known as the Jaynes-Cummings model [4]. The Jaynes-Cummings model has been successfully applied to describe quantum properties of light in such phenomena as vacuum Rabi splitting [5] and quantum Rabi oscillation [6]. In the past decade, theoretical and experimental studies of the matter-light interaction evolve toward the ultrastrong ($0.1 < g < \omega$) and deep strong ($g > \omega$) coupling regime where the contribution of the antiresonant (non-RWA) terms in the coupling Hamiltonian cannot be omitted. In these regimes, the coupling strength $g$ is comparable to, or exceeds, the transition frequency $\varepsilon$ between two energy levels of the quantum system. As a result, the counter-rotating component of an electromagnetic field results in complex dynamics of the field-matter interaction (see, e. g. [7-12]) and makes difficulties for its analytical description.

The strong and ultrastrong regimes of light-matter interaction have been studied both theoretically and experimentally in a variety of solid state systems. The steady-state response of two-level systems, mainly superconducting qubits, under their strong continuous-wave driving have been considered [13,14]. The strong driving of qubits has been investigated in Landau-Zener-Stuckelberg interferometry on quantum dots [15,16] and in hybrid quantum systems composed of a superconducting qubit and a nanomechanical resonator [17] or a superconducting flux qubit and a single nitrogen-vacancy (NV) center [18]. Recently, the ultrastrong regime has been studied in time-resolved experiments with artificial atoms such as superconducting flux [19–21] and charge [22] qubits as well as with a single NV center in diamond [23,24], radiation-dressed states of NV centers [25], nuclear spins [26] and mechanical driving of a single electron spin [27]. Qubit-oscillator systems and their potential for preparing nonclassical states [28] as well as multiphoton quantum Rabi oscillations in circuit QED systems [29] have also been studied in this regime. More recently, the unusual features of the deep strong regime have been reported [30-32].

Usually the qubit's dynamics under strong driving is described within the framework of Floquet theory, where the state of a driven system is expressed in terms of quasienergies and quasienergy states [21]. The presence of the various frequency components in the observed Rabi oscillations and the Bessel-function dependence of the quasienergy difference on the driving strength have been demonstrated [20–22]. Dissipative and decoherence processes limiting the observation of Rabi oscillations under the ultrastrong driving regime have been considered theoretically [33,34]. The long-time dynamics of Rabi model under the ultrastrong atom-cavity coupling for driven-dissipative scenario where the cavity is driven by a monochromatic coherent field has been studied [35]. It is interesting that the ultra- and deep strong driving field induces the transitions between the levels of the two-level system not only at its resonant or near resonant excitation, but also when the frequency of electromagnetic driving field $\omega$ is far away from resonance and exceeds significantly the qubit transition frequency $\varepsilon$ [33,34].

In the present paper, we study the dissipative Rabi model under deep strong far-off-resonant driving ($\omega, g \gg \varepsilon$). An analytical description is realized in the framework of the non-secular perturbation theory based on the Krylov–Bogoliubov–Mitropolsky (KBM) averaging method. In section 2 we consider the dissipative dynamics of the qubit. The steady-state first-order field correlation function and the photon emission spectrum are presented in section 3.

## 2. Dissipative dynamics of qubit

### 2.1. Master equation

The master equation for the qubit interacting with a linearly polarized electromagnetic field is

$$i\frac{\partial \rho}{\partial t} = [H, \rho] + i\Lambda\rho, \tag{1}$$

where

$$H = \varepsilon s^z - 2gs^x \cos\omega t \tag{2}$$

is the Hamiltonian and $\Lambda$ is the relaxation superoperator defined as

$$\Lambda\rho = \frac{\gamma_{21}}{2} D[s^-]\rho + \frac{\gamma_{12}}{2} D[s^+]\rho + \frac{\eta}{2} D[s^z]\rho \tag{3}$$

Here $g$ is the qubit-field coupling strength, $s^{\pm,z}$ are components of the pseudospin operator, describing the qubit state and satisfying the commutation relations: $[s^+, s^-] = 2s^z$, $[s^z, s^\pm] = \pm s^\pm$. In addition, $D[O]\rho = 2O\rho O^+ - O^+O\rho - \rho O^+O$, $\gamma_{21}$ and $\gamma_{12}$ are the rates of photon radiative processes from the excited state $|2\rangle$ of the qubit to its ground state $|1\rangle$ and vice versa, and $\eta$ is the dephasing rate. Note that the strong field-qubit interaction should be taken into account at the microscopic definition of the relaxation superoperator (see, e.g., [33, 36-38]). However, for our purposes, it is sufficient to use the superoperator in its standard form, assuming that the relaxation parameters are defined phenomenologically.

After the canonical transformation $\rho_1 = u^+\rho u$, where $u = \exp(i\frac{2g}{\omega} s^x \sin\omega t)$, the master equation is transformed into

$$i\frac{\partial \rho_1}{\partial t} = [H_1, \rho_1] + i\Lambda_1\rho_1, \tag{4}$$

$$H_1 = u^+ H u - iu^+ \frac{\partial u}{\partial t} = (\varepsilon/2)\left[s^z + (s^+ - s^-)/2\right] f(t) + h.c., \tag{5}$$

$$\Lambda_1\rho_1 = u^+\Lambda u\rho_1 = \frac{\Gamma_\downarrow}{2} D[s^-]\rho_1 + \frac{\Gamma_\uparrow}{2} D[s^+]\rho_1 + \frac{\Gamma_\varphi}{2} D[s^z]\rho_1, \tag{6}$$

where

$$\Gamma_\downarrow = (\gamma_{21}a_+^2 + \gamma_{12}a_-^2 - \eta c^2/4)/2, \ \Gamma_\uparrow = (\gamma_{12}a_+^2 + \gamma_{21}a_-^2 - \eta c^2/4)/2,$$

$$\Gamma_\varphi = \left(\eta d^2 - (\gamma_{12} + \gamma_{21})c^2\right)/2; \tag{7}$$

$a_{\pm} = [1 \pm d]$, $c = \text{Im}(f(t))$, $d = \text{Re}(f(t))$, $f(t) = \exp[ia\sin\omega t]$, and $a = 2g/\omega$.

*2.2. Density-matrix time evolution*

The rapidly oscillating terms in the transformed master equation can be eliminated in the framework of the non-secular perturbation theory by using the KBM averaging method [39]. The description of this method and its applications to studies of the dynamics of two-level systems under mono- and bichromatic driving are presented in Refs. [34,40,41]. In the high-frequency limit, $\varepsilon/\omega \ll 1$, and in the first order of non-secular perturbation theory the Hamiltonian $H_1$ is replaced by its effective counterpart:

$$H_1 \to H_{eff}^{(1)} = \langle H_1 \rangle = \varepsilon J_0(a) s^z \equiv \varepsilon_q s^z. \tag{8}$$

Here the symbol $\langle ... \rangle$ denotes time averaging over the period $2\pi/\omega$ of the rapid oscillations $\exp(im\omega t)$ given by $\langle O(t) \rangle = \dfrac{\omega}{2\pi} \int\limits_0^{2\pi/\omega} O(t)dt$, where $m = \pm 1, \pm 2, ...$, and $O(t)$ is some time-dependent operator. $J_n$ is the Bessel function of the first kind ($n = 0, \pm 1, \pm 2, ...$). The second order of the nonsecular perturbation theory does not yield the contribution in the effective Hamiltonian. The third order correction to the quasienergy $\varepsilon_q$ is small [34] and is neglected here. The relaxation operator calculated in the first non-vanishing approximation is given by

$$\langle \Lambda_1 \rangle \langle \rho_1 \rangle = \frac{\langle \Gamma_\downarrow \rangle}{2} D[s^-]\langle \rho_1 \rangle + \frac{\langle \Gamma_\uparrow \rangle}{2} D[s^+]\langle \rho_1 \rangle + \frac{\langle \Gamma_\varphi \rangle}{2} D[s^z]\langle \rho_1 \rangle, \tag{9}$$

where

$$\langle \Gamma_\downarrow \rangle = \frac{\gamma}{8}\left(3 + 4J_0(a) + J_0(2a)\right) - \frac{\eta}{8}\left(J_0(2a) - 1\right),$$

$$\langle \Gamma_\uparrow \rangle = \frac{\gamma}{8}\left(3 - 4J_0(a) + J_0(2a)\right) - \frac{\eta}{8}\left(J_0(2a) - 1\right), \quad \langle \Gamma_\varphi \rangle = \frac{\gamma + \eta}{2} + \frac{\eta - \gamma}{2} J_0(2a). \tag{10}$$

In the above equations, we denoted $\gamma_{21}$ by $\gamma$ and assumed that at low temperatures $\gamma_{12} \approx 0$. Then, we obtain

$$\frac{i\partial \langle \rho_1 \rangle}{\partial t} = \left[H_{eff}, \langle \rho_1 \rangle\right] + i\langle \Lambda_1 \rangle \langle \rho_1 \rangle. \tag{11}$$

Taking into account that

$$\rho(0) = 1/2 - s^z, \tag{12}$$

the density matrix in the laboratory frame can be written as

$$\rho(t) = \frac{1}{2} + \left(\sigma_0 - (\sigma_0 + 1)e^{-\langle \Gamma_\parallel \rangle t}\right)\left(\cos(a\sin\omega t)s^z - \frac{i}{2}\sin(a\sin\omega t)(s^+ - s^-)\right), \tag{13}$$

where

$$\sigma_0 = -(\langle\Gamma_\downarrow\rangle - \langle\Gamma_\uparrow\rangle)/\langle\Gamma_\parallel\rangle, \tag{14}$$

$$\langle\Gamma_\parallel\rangle = \langle\Gamma_\downarrow\rangle + \langle\Gamma_\uparrow\rangle = \frac{3}{4}\gamma + \frac{1}{4}\eta + \frac{1}{4}(\gamma - \eta)J_0(2a),$$

$$\langle\Gamma_\perp\rangle = (\langle\Gamma_\downarrow\rangle + \langle\Gamma_\uparrow\rangle + \langle\Gamma_\varphi\rangle)/2 = \frac{5}{8}\gamma + \frac{3}{8}\eta + \frac{1}{8}(\eta - \gamma)J_0(2a). \tag{15}$$

We find the population difference of the initial qubit

$$W = tr(s^z \rho(t)) = \Xi(t)\cos(a\sin\omega t) \tag{16}$$

and the dipole moment

$$D = tr(s^+ \rho(t)) = \frac{i}{2}\Xi(t)\sin(a\sin\omega t), \tag{17}$$

where

$$\Xi(t) = \left[\sigma_0 - (\sigma_0 + 1)e^{-\langle\Gamma_\parallel\rangle t}\right]/2 \tag{18}$$

is the relaxation coefficient, and $\sigma_0 = -\gamma J_0(a)/\langle\Gamma_\parallel\rangle$ in accordance to Eq. (14). When the relaxation is ignored ($\langle\Gamma_\parallel\rangle = 0$), the expression (16) for $W$ coincides with the one obtained for the strong quantum field [10,42]. Within the approximation considered here ($\omega, g \gg \varepsilon$), $\varepsilon$ does not enter the expressions for the relaxation parameters and the population difference. In order to applying the KBM method, we must use the values of $a \geq 1.5$.

*2.3. Oscillations in population difference*

It follows from Eq. (13) that the qubit dynamics depends strongly on the parameter $\sigma_0$ which is equal to twice the population difference of the quasienergy states at $t \to \infty$. Due to the presence of $J_0(a)$ in the expression for $\sigma_0$, this parameter oscillates and can be negative, zero or positive (Fig. 1). We observe that the ratio of the rates of energetic relaxation and pure dephasing influence only the amplitude of the variations in $\sigma_0$, but cannot change the sign of $\sigma_0$. As a result, three qualitatively different regimes of the oscillations in the population difference $W$ can be observed. These regimes have been demonstrated in Ref. [34]. We distinguish the collapse and revival of oscillations in the population difference (at $\sigma_0 > 0$), the simple exponential decay to the steady-state oscillations (at $\sigma_0 < 0$), and the exponential vanishing of the oscillations (at $\sigma_0 = 0$).

The function $\cos(a\sin\omega t)$ in Eq. (16) can be evaluated by the following series: $J_0(a) + 2\sum_{n=1}^{\infty} J_{2n}(a)\cos(2n\omega t)$. It means that the oscillations of the population difference under the deep strong far-off-resonant driving occur at even harmonics of the external field. At the same time, the oscillations of the dipole moment (Eq. (17)) occur at odd harmonics because $\sin(a\sin\omega t) = 2\sum_{n=1}^{\infty} J_{2n-1}(a)\sin[(2n-1)\omega t]$.

Fig. 1b depicts the qubit relaxation rate $\langle \Gamma_\| \rangle$ of the population difference versus the normalized driving strength $a = 2g/\omega$. The features of this dependence are determined by the ratio of the rates of energetic relaxation $\gamma$ and pure dephasing $\eta$. The relaxation rate $\langle \Gamma_\| \rangle$ oscillates in accordance with the Bessel-function dependence $J_0(2a)$ for $\gamma < \eta$ or $\gamma > \eta$ and is independent of the driving strength at $\gamma = \eta$. There is a good agreement between the analytical and numerical results presented in Fig. 1b by the solid and dotted lines, respectively.

## 3. Correlation function and emission spectrum of qubit

Now we consider the photon emission by the qubit under its deep strong off-resonant driving. The power spectrum $S(\Omega)$ of the emission is given by the Fourier transform of the first order correlation function $g^{+-}(t,\tau) = tr\left(s^+(t)s^-(t+\tau)\rho(0)\right)$:

$$S(\Omega) = \frac{1}{\pi}\text{Re}\left[\int_0^\infty d\tau e^{i\Omega\tau} \lim_{t\to\infty} g^{+-}(t,\tau)\right]. \tag{19}$$

This function can be split into two parts

$$g^{+-}(t,\tau) = tr\left(s^+\rho(t)\right)tr\left(s^-\rho(t+\tau)\right) + tr\left(\delta s^+(t)\delta s^-(t+\tau)\rho(0)\right), \tag{20}$$

where the dipole moment $s$ is written as a sum of an average dipole moment $\langle s \rangle$ and the instantaneous difference $\delta s$ from its average value: $s = \langle s \rangle + \delta s$. The first term in Eq. (20) is the square of the dipole average moment of Eq. (17) and describes the coherent scattering of radiation. The second term related to fluctuations of the dipole moment decreases with some correlation time and presents the incoherent component of the scattered radiation. Consequently, the emission spectrum from Eqs. (19) and (20) can be written as a sum of "coherent" and "incoherent" parts:

$$S(\Omega) = S_{coh}(\Omega) + S_{inc}(\Omega). \tag{21}$$

Using Eq. (17), we find

$$S_{coh}(\Omega) = \sigma_0^2 \sum_{n=1}^{\infty} J_{2n-1}^2(a) \left[ \delta(\Omega - (2n-1)\omega) + \delta(\Omega + (2n-1)\omega) \right]. \tag{22}$$

The "coherent" part of the spectrum in Eq. (22) consists of a series of narrow delta-like peaks (multiphoton quantum transitions) only at odd harmonics of the exciting field: $\omega$, $3\omega$, ..., $(2n-1)\omega$, where $n = 1, 2, ...$. So, under the deep strong off-resonant driving, the qubit is parametrically excited only at odd harmonics of the external field and coherently re-emits at these frequencies.

The "incoherent" part of the correlation function is calculated by invoking the quantum regression theorem [43] and can be written as

$$S_{inc}(\Omega) = \frac{1}{4\pi}(1 + \sigma_0 J_0(a))\langle \Gamma_\perp \rangle \left\{ \frac{1 + J_0(a)}{(\Omega - \varepsilon_q)^2 + \langle \Gamma_\perp \rangle^2} + \sum_{n=1}^{\infty} J_{2n}(a) \left( \frac{1}{(\Omega - 2n\omega - \varepsilon_q)^2 + \langle \Gamma_\perp \rangle^2} + \frac{1}{(\Omega + 2n\omega - \varepsilon_q)^2 + \langle \Gamma_\perp \rangle^2} \right) \right\}. \tag{23}$$

According to Eq. (23), the "incoherent" part of the multiphoton emission spectrum consists of a series of Lorentzian peaks. A low-frequency peak is centered at the frequency equal to the quasienergy $\Omega^{(0)} = \varepsilon_q$. Other peaks are observed at frequencies $\Omega^{(n)} = 2n\omega + \varepsilon_q$, where $n = 1, 2, ...$. We see that the incoherent photons are emitted at the frequency of the qubit quasienergy $\varepsilon_q$ as well as at even harmonics of the exciting field. These harmonics are shifted towards higher (lower) frequencies by the value of $|\varepsilon_q|$ ($-|\varepsilon_q|$). Relative intensities of the low-frequency and other peaks are given by $1 + J_0(a)$ and $J_{2n}(a)$, respectively. All these spectral lines have the same width $2\langle \Gamma_\perp \rangle$. The central frequencies, widths and relative intensities of the peaks depend on the driving strength. It follows from Eq. (15) that the relaxation rate $\langle \Gamma_\perp \rangle$ oscillates in accordance with the Bessel-function dependence $J_0(2a)$. As a result, the widths of the Lorentzian lines in the "incoherent" emission spectrum oscillate with increasing the driving strength. Like the case of the relaxation rate $\langle \Gamma_\parallel \rangle$, the features of these oscillations are determined by the ratio of the rates of energetic relaxation $\gamma$ and pure dephasing $\eta$. If these rates are equal to each other ($\gamma = \eta$), the relaxation rate of the population difference does not dependent on the driving strength. At $\gamma > \eta$, the variations of $\langle\Gamma_\perp\rangle$ are inverted in comparison with the case $\gamma < \eta$, because the coefficient before the Bessel function in Eq. (15) changes its sign.

For numerical calculations we use the following system of equations for the correlation functions $g^{+-}(t,\tau)$, $g^{+z}(t,\tau) = tr\left(s^+(t)s^z(t+\tau)\rho(0)\right)$, $g^{++}(t,\tau) = tr\left(s^+(t)s^+(t+\tau)\rho(0)\right)$:

$$\frac{\partial}{\partial\tau}g^{+-}(t,\tau) = -i\varepsilon g^{+-}(t,\tau) - ia\omega\cos(\omega\tau)g^{+z}(t,\tau) - \gamma_\perp g^{+-}(t,\tau), \qquad (24)$$

$$\frac{\partial}{\partial\tau}g^{+z}(t,\tau) = (i/2)a\omega\cos(\omega\tau)\left(g^{++}(t,\tau) - g^{+-}(t,\tau)\right) - \gamma_\parallel\left(g^{+z}(t,\tau) + 1/2\right),$$

$$\frac{\partial}{\partial\tau}g^{++}(t,\tau) = i\varepsilon g^{++}(t,\tau) + ia\omega\cos(\omega\tau)g^{+z}(t,\tau) - \gamma_\perp g^{++}(t,\tau)$$

with initial conditions

$$g^{+-}(t,\tau=0) = 1/2 + tr\left(s^z\rho(t)\right),\ g^{++}(t,\tau=0) = 0,\ g^{+z}(t,\tau=0) = -tr\left(s^+\rho(t)\right). \qquad (25)$$

The rates of the transverse and longitudinal relaxations in Eq. (24) are defined as $\gamma_\perp = (\gamma + \eta)/2$ and $\gamma_\parallel = \gamma$. In Fig. 2 we plot the emission spectra $S(\Omega)$ which were calculated numerically using Eqs. (19), (24) и (25) and analytically using the obtained formulas (20) - (23) for different values of the driving strength $a$. These spectra correspond to three different regimes of the qubit's dynamics [34], when the values of $\sigma_0$ are negative (Fig. 2a and b), positive (Fig. 2d and e) and zero (Fig. 2c). The narrow delta-like lines at odd harmonics $(2n-1)\omega$, where $n=1,2,...$, show the "coherent" part of the emission spectrum, $S_{coh}(\Omega)$. The Lorentzian lines at the frequencies $\varepsilon_q$ and $\Omega^{(n)} = 2n\omega + \varepsilon_q$ present the "incoherent" part, $S_{inc}(\Omega)$. There is reasonable agreement between the numerical and analytical results. We illustrate that the driving field can invert some Lorentzian lines as their intensities are proportional to the functions $J_{2n}(a)$, which can change their sign with changing the driving strength $a$. It means that at the certain frequencies the absorption of photons by the qubit can take place instead their emission.

From the numerically calculated emission spectrum we can find the quasienergy $\varepsilon_q$ and compare with its analytical expression $\varepsilon_q = \varepsilon J_0(a)$ in Eq. (8). Fig. 4 shows the quasienergy $\varepsilon_q$ as a function of the driving strength $a$. The analytical results are in a good agreement with the numerical calculations of the frequency shift, $\Omega^{(n)} - 2n\omega = \varepsilon_q$, of the "incoherent" peaks in the emission spectrum relative to even harmonics.

## 4. Conclusions

We have studied the dynamical properties of the dissipative Rabi model under deep strong far-off-resonant driving. We have analytically described the oscillations in the population difference of the qubit and its dipole moment as well as its photon emission spectrum. Moreover, we have found that the decay rate of the time oscillations in the population difference and the width of emission lines are the Bessel-function-like dependences on the driving strength. The multiphoton emission spectrum consists of the coherent (the narrow delta-like lines at odd harmonics of the driving field) and incoherent (the Lorentzian lines at frequencies equal to the quasienergy and the sum of the quasienergy and even harmonics of the driving field) parts. Our results are confirmed by numerical calculations. The discovered features of the Rabi model under deep strong far-off-resonant driving are fundamental and important to the physics of open quantum systems as well as for practical applications. In particular, the predicted regime with the steady-state oscillations can be used for long-time manipulations to quantum information because in this regime the coherent oscillations are not limited by the relaxation times.

**Figures**

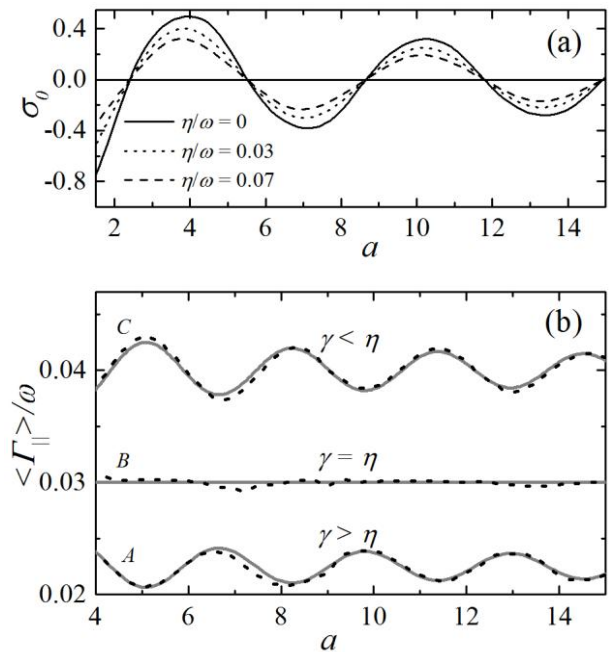

Fig 1. (a) The parameter $\sigma_0$ as a function of the driving strength at the energetic relaxation rate $\gamma/\omega = 0.03$ and three values of the pure dephasing rate $\eta$. (b) The relaxation rate $\langle \Gamma_\parallel \rangle$ of the population difference as a function of the driving strength at the energetic relaxation rate $\gamma/\omega = 0.03$ and the dephasing rate $\eta = 0$ (line A), $\gamma = \eta$ (line B), $\eta/\omega = 0.07$ (line C). The dashed dotted lines present the numerical results.

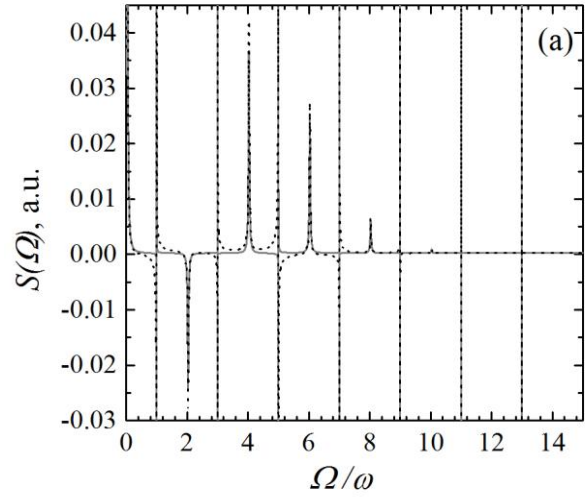
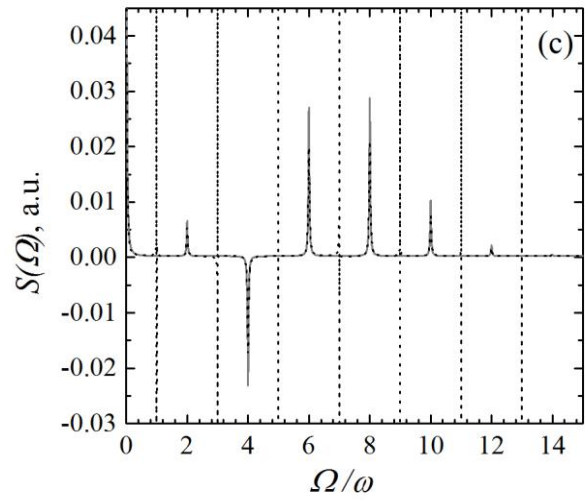
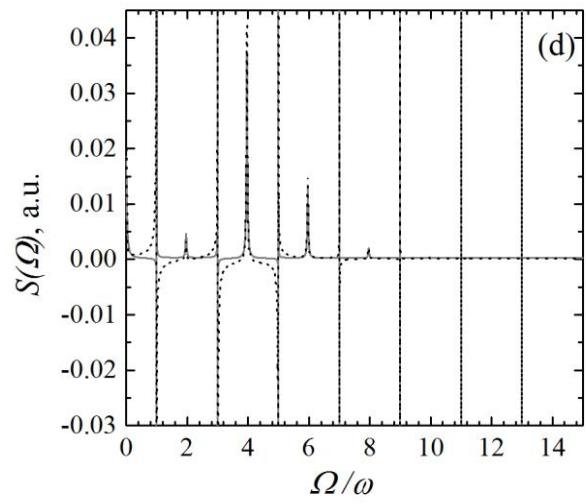

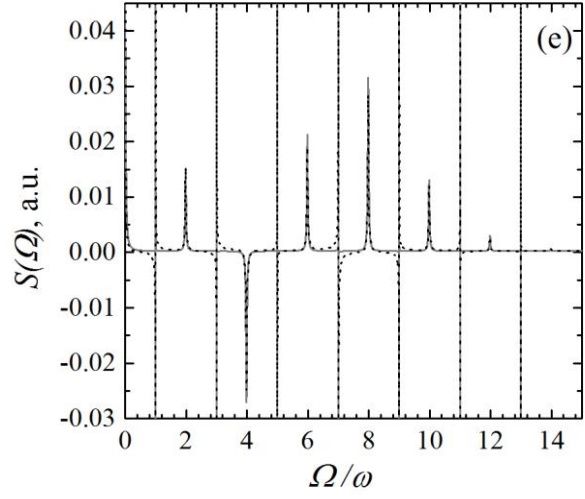

Fig 2. The emission spectra for different values of the driving strength at $\gamma/\omega = 0.03$ and $\eta = 0$. (a) $a = 6.0$, $\sigma_0 = -0.198$ (b) $a = 12.0$, $\sigma_0 = -0.065$ (c) $a = 8.6$, $\sigma_0 = 0$ (d) $a = 5.0$, $\sigma_0 = 0.258$ (e) $a = 9.0$, $\sigma_0 = 0.12$. The dotted lines show the numerical results.

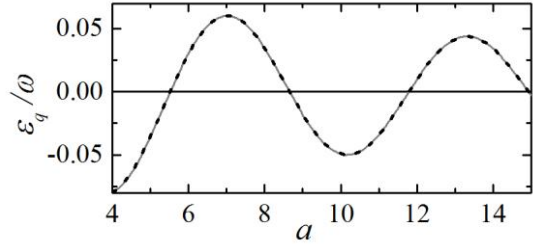

Fig. 3. The quasienergy $\varepsilon_q$ as a function of the driving strength at $\gamma/\omega = 0.03$ and $\eta = 0$. The dashed line presents the numerical results.